\documentstyle[12pt,epsfig]{article}
\input epsf

\large
\newcommand{\non}{\nonumber}
\newcommand{\be}{\begin{eqnarray}}
\newcommand{\ee}{\end{eqnarray}}
\newcommand\del{\partial}
\newcommand{\Dirac}{\rlap {\hspace{-0.5mm} \slash} D}
\newcommand{\mat} {\left ( \begin{array}{cc}}
\newcommand{\emat} { \end{array}\right )}
\newcommand{\matt}{\left ( \begin{array}{ccc}}
\newcommand{\ematt}{\end{array} \right )}
\newcommand{\matf}{\left ( \begin{array}{cccc}}
\newcommand{\ematf}{\end{array} \right )}
\newcommand{\vect}{\left ( \begin{array}{c}}
\newcommand{\evect}{\end{array} \right )}

\begin{document}
\setlength{\baselineskip}{17pt}
\pagestyle{empty}
\vfill
\eject
\begin{flushright}
SUNY-NTG-00/12
\end{flushright}

\vskip 2.0cm
\centerline{\Large \bf  The Replica Limit of Unitary Matrix Integrals}

\vskip 1.2cm
\centerline{ D. Dalmazi\footnote{On leave from UNESP 
(Guaratinguet\'{a}-Brazil)   } and J.J.M. Verbaarschot}
\vskip 0.2cm
\centerline{\it Department of Physics and Astronomy, SUNY, 
Stony Brook, New York 11794}
\vskip 0.2cm
\centerline{\tt dalmazi@tonic.physics.sunysb.edu,\, 
verbaarschot@nuclear.physics.sunysb.edu}
\vskip 1.5cm

\centerline{\bf Abstract}
We investigate the replica trick for the microscopic spectral 
density, $\rho_s(x)$, of the Euclidean QCD Dirac operator. 
Our starting point is the low-energy limit of the QCD 
partition function for $n$ fermionic flavors (or replicas)
in the sector of topological charge $\nu$. In the domain of the smallest 
eigenvalues, this partition function is simply given by 
a $U(n)$ unitary matrix 
integral. We show that the asymptotic behavior  
of $\rho_s(x)$ for $x \to \infty$ is obtained from the $n\to 0$ limit of this 
integral. The smooth contributions to this series are obtained 
from an expansion  about the 
replica symmetric saddle-point, whereas the oscillatory terms 
follow from an expansion about a saddle-point that breaks the replica 
symmetry. 
~For $\nu =0$ we recover the small-$x$ logarithmic singularity of the 
resolvent by means of the replica trick.
~For half integer $\nu$, when the saddle point expansion of the
$U(n)$ integral terminates, 
the replica trick reproduces the exact analytical result. In all 
other cases only an asymptotic series that does not uniquely determine the 
microscopic spectral density is obtained. We argue that bosonic replicas 
fail to reproduce the microscopic spectral density. 
In all cases, the exact answer is obtained naturally by means of the
supersymmetric method.

 
\vfill
\noindent

\eject
\pagestyle{plain}

\section{ Introduction}
In disordered systems, the ensemble average of the logarithm of the
partition function cannot be evaluated directly in most cases. Two 
widely used methods, the replica trick \cite{EA}
and the supersymmetric method \cite{Brezin,Efetov,VWZ}, 
have been proposed to circumvent this problem. In the replica trick
the ensemble  average of the logarithm of the partition function is written as
\be
\langle \log Z\rangle  = \lim_{n \to 0} \langle \frac {Z^n -1}n \rangle,
\ee
i.e., $\log Z$ is calculated by the analytic continuation of
the  $n$-dependence of $n$ replicated partition functions to $n \to 0$. 
If $Z$ is given by a fermion determinant, 
$Z^n$ can be written as an integral over
$n$ replicated Grassmann fields, and therefore,
this method is known as the fermionic replica trick. Alternatively,
$\log Z$ can be written as
\be
 \langle \log Z\rangle  = \lim_{n \to 0} \langle \frac {1-Z^{-n}}n \rangle.
\ee
In this case, a determintal $Z$ can be expressed as
an integral over $n$ replicated complex  fields and this limit
referred to as the bosonic replica trick.
Both for the fermionic and for the bosonic replica trick, the average
partition function can mapped onto a non-linear $\sigma$-model which
is amenable to a saddle-point expansion.

In the supersymmetric method the disorder   
average is performed for the ratio
\be
\left \langle \partial_J \log {Z(J)}  \right \rangle_{J=0} =
\left .\partial_J 
\left \langle \frac{Z(J)}{Z(J=0)} \right \rangle\right |_{J=0}.
\ee  
If $Z(J)$ is given by a determinant, the
numerator can be expressed as a fermionic integral whereas the 
denominator can be written as a bosonic integral. ~For $J=0$ the
partition function is thus invariant with respect to superunitary
transformations that mix the fermionic and bosonic fields. Based on
this symmetry, the supersymmetric partition function can be mapped
onto a supersymmetric non-linear $\sigma-$model \cite{Efetov,VWZ} which
can be used to derive exact analytical expressions for spectral correlation 
functions.  

he replica trick has two obvious advantages. First, it can be applied
to  cases where $Z$ cannot be expressed as a determinant
(as for example is the case in the theory of spin-glasses \cite{SK}), 
and, second,
it is possible to calculate the logarithm of the partition function.
One disadvantage of this method is that, in order to take the
replica limit, the $n$ dependence of the replicated partition
function has to be known analytically. Therefore, the application of 
replica trick is limited to perturbative expansions of the partition function.
A much more serious problem of this method is that the continuation of
the $n$-dependence to $n\to 0$ is not unique.
~For example,  a term of the form $\sim \sin n\pi$ contributing to $Z^n$
vanishes for integer $n$
but gives rise to a nonzero result in the replica limit.
The failure of the replica trick was first noticed in the theory
of spin glasses \cite{SK} where a replica symmetric minimum of the free energy
resulted in a negative entropy
\cite{Mezard}. However, in that case the problems
could be resolved by means of an elaborate scheme of replica symmetry 
breaking \cite{Mezard}. More recently, the replica trick was criticized
because of its failure to reproduce the oscillatory contributions 
to random matrix correlation
functions \cite{VZ}. Notice however that the non-oscillatory contributions
to the two-point function were reproduced correctly \cite{EJ,VZR}. 
Another example for which the replica trick may be problematic are 
nonhermitian Random Matrix Theories with eigenvalues scattered
in the complex plane \cite{misha}, but we will not study such theories in
this article. 

The advantage of the supersymmetric method is that it is possible
to derive non-perturbative analytical results. This has been shown
convincingly for the calculation of spectral correlation
functions of random matrix-ensembles \cite{Efetov,VWZ,VZ,OTV,DOTV}.  
A disadvantage is that it 
requires some familiarity with supermathematics, but for perturbative
expansions this method is no more complicated than the replica trick.
In the example that will be discussed in this article, the 
exact super-symmetric calculation is actually much simpler than the
perturbative replica calculation.
A second disadvantage is, that because the average partition function is
normalized to unity, one does not have access to the average free energy.    

Recently, the replica trick was revived in an article by Kamenev and
M\'ezard \cite{km}. 
They found that, in order to reproduce the oscillatory terms in
spectral two-point correlation function,
saddle points with broken replica symmetry had to be taken into
account \cite{km}. 
~For the Gaussian Unitary Ensemble they found
the exact analytical result. As explained in an article by
Zirnbauer \cite{zirn}, the reason for this miracle is a consequence of the
Duistermaat-Heckman theorem \cite{duister}
which is applicable to the $\sigma-$model for the two-point
function of the Gaussian Unitary Ensemble. This theorem states the
conditions under which an integral is localized on its critical points
so that a saddle-point approximation becomes exact.
~For the Gaussian Orthogonal Ensemble and the Gaussian Symplectic Ensemble
the replica trick could only reproduce the asymptotic expansion 
of the two-point spectral correlation function for large energy differences
\cite{lerner,km2}. We remind the reader that, unless we know the analytical
properties of a function in the complex plane, it cannot be reproduced
from its asymptotic series. A version of the replica trick that does
not rely on the non-linear $\sigma$-model but instead on orthogonal
polynomials was shown to reproduce the exact correlation functions for
all Gaussian ensembles \cite{kanzieper}. In this article we will not
discuss this variant of the replica trick which is not an alternative
to the orthogonal polynomial method.

To investigate the replica trick we have chosen the microscopic spectral
density of the QCD Dirac operator which is defined as the spectral
density near zero on the scale of the average level spacing.   
The reason is three-fold.
i) The $n$-fold replicated partition function is the QCD partition
function with $n$ flavors. Its low-energy limit, relevant for the 
microscopic spectral density, is known analytically.
Because of spontaneous breaking of chiral symmetry in QCD, it is given
by a partition function of weakly interacting Goldstone bosons and, on
the scale of the average level spacing, it can be reduced to a
unitary matrix integral which can be evaluated analytically for 
any number of flavors \cite{Brower,LS,mironov}. 
This, so called finite volume chiral partition
function has been investigated in great detail, also in the 
context of one-link integrals in lattice QCD \cite{Brower,GW}.
ii) Because the eigenvalues occur in pairs $\pm \lambda$, the level
repulsion of the eigenvalues leads to a nontrivial oscillatory 
behavior of the microscopic spectral density.
iii) The low energy partition function can be derived as a function
of two integer valued parameters, the topological charge $\nu$ 
and the number of physical
flavors $N_f$ (in this article only $N_f =0$ will be discussed) and  can be
trivially continued to non-integer $\nu$. ~For half-integer $\nu$,
the saddle point expansion of the $U(n)$-integrals terminates for finite
positive integer values of $n$. This is closely related to 
Duistermaat-Heckman localization, where the leading order saddle point
approximation is exact such as for 
the  $\sigma-$model with $n$ fermionic replicas 
of the two-point function of the Gaussian Unitary
Ensemble \cite{zirn}.
Based on recent work \cite{km,zirn}, we expect that the replica trick 
with replica-symmetry breaking  gives the exact result in this case.

The replica limit of the finite volume chiral partition function
was first studied in \cite{kim}. It was found that the
asymptotic expansion of the 
valence quark mass dependence
of the chiral condensate in the microscopic region 
(i.e. for large valence quark masses in units of the average eigenvalue
spacing) was reproduced by
the replica trick. The small mass expansion was obtained
up to logarithmic singularities which are essential for the calculation
of the spectral density.
Recently, the partially quenched supersymmetric chiral Lagrangian
\cite{Golterman,OTV,DOTV,DT}
was formulated in terms of the replica trick \cite{poul}.

In this article we analyze the oscillatory
contributions to the microscopic spectral density
in the framework of the replica trick. In section 2 
we discuss the chiral symmetries for bosonic replicas and introduce the
supersymmetric quenched low energy chiral partition function. 
The calculation of the
microscopic spectral density by means of the supersymmetric method
is given in section 3. This calculation illustrates
 that the compact/non-compact
structure of the final result for the resolvent appears naturally in
the supersymmetric method. In section 4.1 we derive the large-mass
asymptotic behavior of the microscopic spectral density by means
of the fermionic replica trick. The small-mass behavior is discussed
in section 4.2. Bosonic replicas are discussed in section 5 and
concluding remarks are made in section 6. In the Appendix  we the derive
trace correlators necessary for the replica calculation to 
fourth order in the inverse mass. 
 
\section{Chiral Symmetry}

In this section we discuss chiral symmetry for bosonic and fermionic replicas,
as well as for the supersymmetric partition function. 
To illustrate the difference in flavor symmetries
between bosonic and fermionic replicas we present a detailed discussion
for the case of one flavor or replica. 

The question we wish to address in this article is whether the spectrum
of the QCD Dirac operator in the sector of topological charge $\nu$
can be obtained from the QCD partition function with $n$ additional
replica flavors with quark mass $z$. This partition function is given by
\be
{ Z}_{\nu}^{(N_{f}+n)}(z) ~=~ 
\int\! [dA]_{\nu}
~{\det}^n(i\Dirac + z)\prod_{f=1}^{N_{f}}
{\det}(i\Dirac + m_f) ~e^{-S_{YM}[A]} ~\label{z1}, 
\ee
where $m_1, \cdots, m_{N_f}$ are the usual quark masses. The integral
is over all gauge fields in the sector of topological charge $\nu$ 
(which is chosen positive in this article) and
is weighted by the Yang-Mills action.
The replica limit of the resolvent or the chiral condensate
is defined by
\be
\Sigma(z) \equiv 
\frac 1{V_4} {\rm Tr} \left \langle \frac 1{z+i\Dirac}
\right \rangle \, =\,  \lim_{n\to 0}
\frac{1}{n}\frac 1{V_4} \frac{\partial}{\partial z}
\ln { Z}_{\nu}^{(N_{f}+n)}(z) ~.\label{cond}
\ee
Here, $V_4$ is the Euclidean 4-volume.
The spectral density, which in terms of the eigenvalues $i\lambda_k$ of the
Dirac operator is defined by
\be
\rho(\lambda) = \sum_k \delta(\lambda -\lambda_k),
\ee
follows from the discontinuity across the imaginary axis
\be
\frac{\rho(\lambda)}{V_4} = \frac 1{2\pi} [ \Sigma(i\lambda + \epsilon)
-\Sigma(i\lambda - \epsilon)].
\ee
Below we only consider the dimensionless ratio
\be
\frac {\Sigma(z)}{\Sigma_0} &=& \frac 1{V_4\Sigma_0}
\int_{-\infty}^\infty d\lambda \frac{\rho(\lambda)}
{ z + i\lambda},\nonumber \\
&=& \int_{-\infty}^\infty du \frac 1{V_4\Sigma_0}{\rho(\frac u{V_4\Sigma_0})}
\frac 1{V_4\Sigma_0 z + iu},
\ee
as a function of the microscopic variable $V_4\Sigma_0 z$. 
The chiral
condensate, $\Sigma_0$, is defined as the limit of $\Sigma(z)$ 
for $z$ close to $z=0$ but many level spacings away from the center
of the spectrum.  
As one can see from the second equality, it can be expressed as an integral
over the microscopic spectral density defined by \cite{SVR}
\be
\rho_s(u) = \frac 1{V_4\Sigma_0}{\rho(\frac u{V_4\Sigma_0})}.
\label{micro}
\ee

~For spontaneously broken chiral symmetry, the spacing of the eigenvalues
is given by $\pi/\Sigma_0 V_4$ so that (\ref{micro}) is stable in $V_4$
and can be calculated in the thermodynamic limit. 

Because the resolvent $\Sigma(z)$ has a cut along the imaginary axis
it is sometimes more convenient  to use the relation
$G(i\lambda - \epsilon)= -G(-i\lambda +\epsilon)$ to rewrite
the discontinuity as
\be
\frac{\rho(\lambda)}{V_4} = \frac 1{2\pi} [ \Sigma(i\lambda + \epsilon) 
+\Sigma(-i\lambda + \epsilon)].
\label{discrho}
\ee
We thus only need to calculate the resolvent  in the half-plane
${\rm Re}(z) > 0$.
If the argument $\Sigma(z)$ represents  the microscopic variable, this
relation gives us the microscopic spectral density.

The reason of working with the partition function (\ref{z1})
is that in the phase of spontaneously broken symmetry its low
energy limit in entirely determined by chiral symmetry and is
a partition function of weakly interacting Goldstone modes (or pions).
We are interested in the kinematical domain \cite{LS}
\be
1/m_{\pi} \gg V^{1/4}_4\gg 1/\Lambda_{QCD}.
\label{range}
\ee
Because $V^{1/4}_4 \gg 1/\Lambda_{QCD}$, only the Goldstone modes
contribute to the mass-dependence of the partition function \cite{GL}.
~For quark masses for which the Compton wavelength of the Goldstone modes
is much larger than the size of the box ( $1/m_{\pi} \gg V^{1/4}_4$), the
kinetic term of the chiral Lagrangian can be ignored, 
and only the constant fields contribute to the mass dependence of the
low-energy partition function. Therefore, as we will see next, in the
domain (\ref{range}) the QCD parition function can be reduced to
a unitary matrix integral. 

In QCD with $n$ fermionic flavors the chiral symmetry group is given
by $U_V(n) \times U_A(n)$. A $U_A(1)$ subgroup of the axial symmetry
group is broken by the anomaly. The remaining
axial symmetry group is broken spontaneously
by the formation of a nonzero chiral condensate and the vector
symmetry group, $U_V(n)$, remains unbroken. 
The Goldstone manifold is thus given by the axial group $SU_A(n)$. The low
energy limit of the QCD partition function
is uniquely fixed from the requirement
that its transformation properties under the chiral symmetry group are
the same as for full QCD. Taking into account the anomaly, one finds,
for $n$ flavors all with mass $m$, in the sector of topological charge $\nu$,
the low-energy finite volume partition function \cite{LS}
\be
Z_{\nu}^{(n)}(x)= \int_{U\in U(n)} 
dU\,  \left(\det U\right)^{\nu} e^{ 
\frac x2 {\rm Tr }(U+U^{-1})},\label{zcomp}
\ee
where $x\equiv m V_4 \Sigma_0$.  
This partition function is valid in the range (\ref{range}).

In the domain (\ref{range}), the low energy partition function can also
be obtained from a Random Matrix Theory with the global symmetries of the QCD
partition function \cite{SVR}. In this theory, the matrix elements of
the Euclidean Dirac operator are replaced by independently distributed
Gaussian random variables.
In the sector of topological charge $\nu$, the Dirac matrix
is thus has given by
\be
D  = \mat 0 & iW \\ iW^\dagger &0 \emat,
\label{Dirac}
\ee 
where $W$ is an $n\times (n+\nu)$ matrix, and the integration over
the gauge fields is replaced by an integration  over the probability
distribution of the matrix elements. If $W$ is complex 
and the probability distribution is a function of  traces
${\rm Tr}( W^\dagger W)^p$, this ensemble is known as the 
chiral Unitary Ensemble (chUE) or the chiral
Gaussian Unitary Ensemble (chGUE) if the distribution of the matrix
elements is Gaussian.
As was shown in  \cite{ADMN}, the statistical properties of the 
smallest eigenvalues of $D$
do not depend on the details of the probability
distribution of the matrix elements.

In order to analyze the chiral symmetry for bosonic replicas
let us first discuss the flavor symmetries for one flavor.
The fermion determinant
that occurs in the QCD partition function can be written as
a integral over Grassmann variables.
\be
{\det(D+m)} = \int d\bar 
\chi d\chi e^{-\int d^4x \bar\chi(D+m) \chi}.
\label{zferm}
\ee
The inverse determinant can be written as an integral over
bosonic integration variables
\be
\frac 1{\det(D+m)} = \int d\phi^* d\phi e^{-\int d^4x \phi^*(D+m) \phi}.
\label{zbos}
\ee 
In a well-defined theory the functional integral has to be convergent. 
This is automatically
the case for the Grassmann integration in (\ref{zferm}), 
but the bosonic integrals 
in (\ref{zbos}) is only convergent for 
positive $m$. The symmetries of the partition function should be
compatible with these convergence requirements. 
In particular, $\phi^*$ should be identified with
the complex conjugate of $\phi$, and not as an 
 independent integration variable such as the  fermionic
variables  $\bar\chi $ and $\chi$.  

If we decompose the spinors according
to the block structure of the Dirac operator the vector symmetry of
the massive theory with one fermionic replica (\ref{zferm})
is given by
\be
\vect \chi_1 \\ \chi_2 \evect \rightarrow e^{i\theta} 
\vect \chi_1 \\ \chi_2 \evect, \qquad
\vect \bar\chi_1 \\ \bar\chi_2 \evect \rightarrow e^{-i\theta} 
\vect \bar\chi_1 \\ \bar\chi_2 \evect ,
\ee
and the axial $U(1)$ symmetry of the massless theory can be written as
\be
\vect \chi_1 \\ \chi_2 \evect \rightarrow 
\vect e^{i\theta} \chi_1 \\ e^{-i\theta} \chi_2 \evect, \qquad
\vect \bar\chi_1 \\ \bar\chi_2 \evect \rightarrow  
\vect e^{i\theta}\bar\chi_1 \\ e^{-i\theta}\bar\chi_2 \evect .
\label{axial}
\ee
The $U(1)$ vector symmetry of the massive bosonic theory (\ref{zbos}) 
is the same as for the fermionic theory
\be
\vect \phi_1 \\ \phi_2 \evect \rightarrow e^{i\theta} 
\vect \phi_1 \\ \phi_2 \evect, \qquad
\vect \phi_1^* \\ \phi_2^* \evect \rightarrow e^{-i\theta} 
\vect \phi_1^* \\ \phi_2^* \evect .
\ee
This transformation does not affect the complex conjugation properties
of $\phi$. However, the axial transformation (\ref{axial}) 
applied to the bosonic fields affects their
complex conjugation properties. In this case the axial transformation
that is compatible with the convergence of the bosonic integral is
given by
\be
 \vect \phi_1 \\ \phi_2 \evect \rightarrow 
\vect e^{s}\phi_1 \\ e^{-s}\phi_2 \evect, \qquad
\vect \phi_1^* \\ \phi_2^* \evect \rightarrow  
\vect e^{s}\phi_1^* \\ e^{-s}\phi_2^* \evect .\label{gl/u}
\ee
The axial symmetry group is therefore not $U(1)$ but instead $Gl(1)/U(1)$.
Of course, this axial transformation is also a symmetry of the 
fermionic partition function.

~For  $n$ bosonic flavors the vector symmetry is $U(n)$ whereas
  the axial symmetry is given by
\be
 \vect \phi_1 \\ \phi_2\evect \rightarrow 
\vect e^{H}\phi_1 \\ e^{-H}\phi_2\evect,\qquad
\vect \phi_1^* \\ \phi_2^* \evect \rightarrow  
\vect e^{H}\phi_1^* \\ e^{-H}\phi_2^* \evect,
\label{axialH}
\ee
with the matrix $H$ containing only real elements. The axial 
symmetry group is
thus given by the coset $Gl(n) /U(n)$. An explicit parameterization of
this coset is given by  $AA^\dagger$ with $ A\in Gl(n)$.
We expect that the axial  symmetry of the bosonic partition function 
\be
{ Z}_{\nu}^{(-n)} =\left \langle \frac 1{{\det}^n (D +m)} 
\right \rangle_\nu
\ee
is  broken in the same way as in the fermionic case
 with  a $Gl(1)/U(1)$ coset  broken explicitly by the anomaly
and the remaining part of the coset broken spontaneously by the chiral
condensate. In absence of explicit symmetry breaking, the
Goldstone manifold is thus given by $Gl(n) /U(n)$. 

The mass term introduced according to
\be
\sum_{k,l=1}^n \phi_1^{*\,k} M_{kl}\phi_1^{l}
+\phi_2^{*\,k} M_{kl}^\dagger \phi_2^{l}
\label{massterm}
\ee    
is invariant under the axial transformation (\ref{axialH})
provided that the mass matrix is transformed at the same time as
\be 
M \rightarrow e^{-H} M e^{-H}, \qquad M^\dagger 
\rightarrow e^{H} M^\dagger e^{H}.
\ee
The bosonic partition function in the sector of topological charge $\nu$ 
transforms covariantly  with (\ref{axialH})
\be
{ Z}_{\nu}^{(-n)}\rightarrow \det (e^{2\nu H}) { Z}_{\nu}^{(-n)}.
\ee
The low energy limit of bosonic partition function 
is uniquely fixed by the requirement that its transformation properties 
are the same as of ${ Z}_{\nu}^{(-n)}$.  In the sector of topological
charge $\nu$ it is given by
\be
Z_{\nu}^{(-n)}= \int_{U\in Gl(n)/U(n)} 
dU\,  \left(\det U\right)^{\nu} e^{ 
\frac{\Sigma_0 V_4}2 {\rm Tr }(MU+M^\dagger U^{-1})},
\ee
The measure $dU$ is the invariant Haar measure.
Below we consider only the case with
a diagonal mass matrix with all nonzero matrix elements equal to $m$ 
and use the definition
$x = mV_4 \Sigma_0$. This partition function is valid in the kinematical
domain (\ref{range}) 
where constant Goldstone fields are the only relevant degrees of freedom.

Since fermionic integrals are always convergent the partition function
is invariant under both compact ($U(n)$) and non-compact 
($Gl(n)/U(n)$) axial transformations. 
However, the small mass behavior of the 
non-compact partition function is singular because of the volume of
the non-compact group diverges. This is not the case for the fermionic
partition function and therefore this $Gl(n)/U(n)$ is not an admissible
parameterization of the Goldstone manifold. The correct parameterization is
given by the compact manifold $U(n)$. At a more technical level, this
follows from the fact that the transformations that lead to
the non-compact integral are only legitimate for bosonic quarks. ~For fermionic
quarks one necessarily finds a compact effective partition function. 

In the supersymmetric method  the generating
function of the quenched resolvent 
is given by
\be
Z_\nu(z,J) = \left 
\langle \frac{\det(\Dirac+z+J)}{\det(\Dirac+z)} \right\rangle_\nu.
\label{zgen}
\ee
It  can be written as a superintegral
\be
Z_\nu(M,M^\dagger)\! =\! 
\left \langle\int d\phi d\phi^* d\chi d\bar\chi 
\exp \! \left [\! \phi^*\Dirac \phi + 
\bar \chi \Dirac\chi
+ \vect \phi_1^* \\ \bar\chi_1 \evect \!  M \! \vect \phi_1 \\ \chi_1\evect
+ \vect \phi_2^* \\ \bar\chi_2 \evect \! M^\dagger \! \vect \phi_2 \\ \chi_2
\evect \right  ] \! \right\rangle_\nu\! .\nonumber \\
\label{zgen1}
\ee
Both $M$ and $M^\dagger$ are given by ${\rm diag}(z+J,z)$, but in order
to study the transformation properties of the partition function, we
keep them as general matrices.

With spontaneously broken axial symmetry, the bosonic part
of the Goldstone manifold is $U(1) \times 
Gl(1)/U(1)$. Because, the
partition function is invariant under super-unitary transformations, 
the full symmetry group is given by the maximum Riemannian submanifold
of $Gl(1|1)$ \cite{class,OTV,DOTV} (we will denote this manifold 
by ($\hat {Gl}(1|1)$). 
This manifold
can be parameterized as
\be
U= \mat e^{i\theta} & \alpha \\ \beta &e^s \emat.
\ee
~For zero topological charge
the generating function (\ref{zgen}) is invariant under $Gl_R(1|1)
\times Gl_L(1|1)$ if at the same time the mass matrix is transformed as 
\be
M \rightarrow U_R^{-1} M U_L, \qquad M^\dagger \rightarrow 
U_L^{-1} M^\dagger U_R.
\label{mtrans}
\ee
~For nonzero values of $\nu$ the generating function (\ref{zgen}) 
is not invariant  under (\ref{mtrans}) but transforms
according to
\be
Z_\nu(z,J) \rightarrow {\rm Sdet}^{\nu}(U_R^{-1} U_L) Z_\nu(z,J).
\ee
The low-energy partition function is obtained from the requirement 
that it should have the same transformation properties as the
QCD partition function (\ref{zgen}). In the sector
of topological charge $\nu$, it is given by
\be
Z_\nu(z,J) = \int_{U \in \hat {Gl}(1|1)} dU {\rm Sdet}^\nu (U)
e^{\frac {\Sigma_0 V_4}2 {\rm Str}(M U + M^\dagger U^{-1})}.
\label{superzdef}
\ee
The integration is over the Haar measure of $\hat {Gl}(1|1)$,
Below, we only the consider the case of a diagonal mass matrix
with  both $M$ and $M^\dagger$ equal
to ${\rm diag}(z+J,z)$.

An amusing observation is that 
 the topological charge in QCD partition function
is discrete,  whereas the number of
flavors with equal mass, thought of as the power of the fermion
determinant 
is a continuous parameter. In the low energy partition function
it is just the other way round. 

\section{ Supersymmetric Calculation of the Resolvent}

In this section we calculate the super-integrals in (\ref{superz}) to
obtain analytical expression for the resolvent and
the microscopic spectral density. ~For integer $\nu$, this calculation
was also presented in \cite{OTV,DOTV}.
To simplify the explicit calculations we will choose
the parameterization 
\be
U = \mat e^{i\theta} & 0   \\ 0 &e^s \emat
\exp \mat 0 & \alpha \\ \beta & 0 \emat.
\ee
 In this case the Haar measure is simply given by
\be
dU = d\theta ds d\alpha d\beta.
\ee
This results in the partition function
\be
Z(z,z+J) = \frac 1{2\pi}
\int_{-\infty}^\infty ds \int_{C_c} 
d\theta d\alpha d\beta e^{i\nu\theta-\nu s} 
\exp {\rm Str} M
\mat (1+\frac{\alpha \beta}2)\cos \theta & \alpha(e^s - e^{-i\theta})\\
\beta(e^{i\theta} - e^{-s} ) & (1-\frac {\alpha \beta}2 ) \cosh s \emat,
\nonumber \\
\label{superz}
\ee
Where $z$ and $J$ are now microscopic variables, i.e. they are expressed
in units of $1/V_4\Sigma_0$.
The integration over $s$ is over the complete real axis. ~For integer $\nu$
the integration over $\theta$ is over the interval $[-\pi,\pi]$. 
~For non-integer
$\nu$ the translational invariance of the $\theta$-integral is lost. It
is recovered by extending the integration contour to include
the intervals $\langle -\pi+i\infty, -\pi]$ and $[\pi, \pi+i\infty\rangle$.
A picture of this integration contour, denoted by $C_{c}$, is shown in 
Fig. 1.  

\vspace*{2cm}
\begin{figure}[ht!]  
\begin{center}
\epsfig{figure=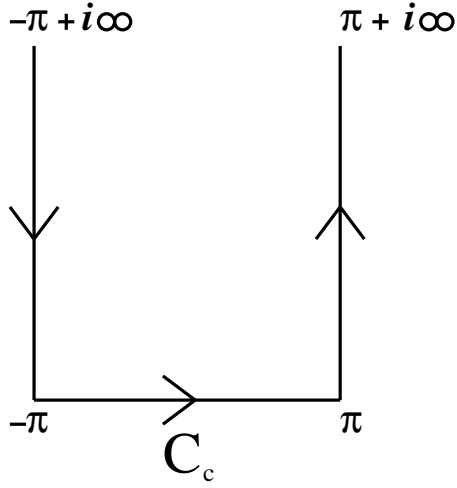,width=60mm}
\caption{Contour for the $\theta$ integration.}
\label{callisto1}
\end{center}
\end{figure} 

After performing the Grassmann integrations the partition function 
reduces to
\be
Z(z,J) &=&\frac 1{2\pi}
\int_{-\infty}^\infty ds \int_{C_c}d\theta e^{i\nu\theta-\nu s} 
({(z+J)} \cos \theta + z \cosh s ) 
e^{{(z+J)} \cos \theta -  z \cosh s} \nonumber \\
&=& 
(z+J)K_\nu(z)I_{\nu+1}(z+J) + zK_{\nu+1}(z) I_{\nu}(z+J).
\label{zfinal}
\ee
The normalization of the partition function according to
$Z(z,J=0)= 1$ follows from the
Wronskian identity  $zK_\nu(z)I_{\nu+1}(z) + zK_{\nu+1}(z) I_\nu(z)=1$.
The resolvent obtained by diferentiation with respect to $J$ is,
after using some identities for Bessel functions, given by
\be
\Sigma(z) = z(K_\nu(z)I_{\nu}(z) +K_{\nu-1}(z) I_{\nu+1}(z))
+\frac \nu z.
\label{valsup}
\ee
The microscopic spectral density then follows from the discontinuity across
the imaginary axis according to (\ref{discrho}),
\be
\rho_s(x) &=& \frac 1{2\pi} 
[\Sigma(i x +\epsilon) +\Sigma(-ix +\epsilon)] \nonumber \\
&=&
\frac x2(J_\nu(x)^2 - J_{\nu+1}(x)J_{\nu-1}(x)) +\nu \delta(x).
\label{micro2}
\ee
The last term is the contribution from the $\nu$ zero modes.
~For integer values of $\nu$ the integration contour $C_c$ can be replaced
by the segment $[-\pi,\pi]$. The restriction of the integration
over $\theta$ to this segment for non-integer values of $\nu$ 
would have resulted in the wrong answer. This can be seen from the
following representation of modified  Bessel functions,
\be
I_\nu(z) =\frac 1{2\pi} \int_0^{2\pi} e^{z\cos\theta} e^{i\nu\theta} d\theta
- \frac {\sin\nu \pi}{\pi} \int_0^\infty e^{-z\cosh s - \nu s} ds.
\ee
The microscopic spectral density and the expression for $\Sigma(z)$ were
first obtained from chiral Random Matrix Theory by means of
the orthogonal polynomial method \cite{VZm,Vplb} and can also
be obtained by means of the supersymmetric method \cite{Simons}. I
n that case
the derivation, starting from the joint probability distribution of
the eigenvalues, is also correct for non-integer values of 
$\nu$ and is given by
the expressions (\ref{valsup}) and (\ref{micro2}).

The asymptotic expansion of the imaginary part of $\Sigma(z)$ is given by
\be
i{\rm Im}\Sigma(z) 
&=&  -\frac{i(-1)^{\nu}e^{-2z}}{2z} + 
\frac{-i(4\nu^2-1)(-1)^{\nu}e^{-2 z}}{8z^2}
-\frac{i(-1)^{\nu}e^{-2z}(4\nu^2-1)(4\nu^2-9)}
{64z^3}\nonumber\\ 
&+& \frac{i(4\nu^2-1)(4\nu^2-9)(19-4\nu^2)
(-1)^{\nu}e^{-2z}}{3\times 256 z^4},
\label{cond4p}
\ee
and for the asymptotic expansion of the spectral density we find
\be
\rho_{s\,\nu}(x)\!  &=& \!
\frac 1{\pi }\left\lbrack 1 - \frac{ \cos (2x-\pi\nu)}{2x} + 
\frac{1-4\nu^2}{8x^2} (1-\sin (2x -\pi\nu)) +
(4\nu^2-1)(4\nu^2-9)\frac{\cos (2x-\pi\nu)}
             {64 x^3}\right .
\nonumber \\
&+& \left .(4\nu^2-1)(4\nu^2-9)\frac{( -6 + (19-4\nu^2) \sin (2x-\pi\nu ))}
                    {x^4 2^7 3!} + \cdots \right\rbrack .
\label{rho4}
\ee
The asymptotic expansion of the partition function, the resolvent, and
the spectral density  
terminates for half-integer values of $\nu$. ~For example, for 
$\nu =1/2$ only
one oscillating term in (\ref{rho4}) is nonvanishing suggesting 
that it can be obtained from
a leading order saddle point approximation.

Since for half-integer $\nu$ the integral is localized on the critical points,
we expect that the asymptotic expansion generated by the replica trick
reproduces the exact answer. ~For other values of $\nu$ we expect 
that the replica trick  reproduces
the asymptotic series to all orders in $1/x$.

%For $\nu=3/2$ only one additional term contributes to the
%partition function, but this is no longer true for higher order terms.%
%
%I can be shown that the the $\nu$-dependence of the $O(1/x)$ term
%is given by $(-1)^\nu$. The replica calculation is independent for
%$\nu$ and since it is exact for $\nu = 1/2 $, it is exact for all $\nu$.
%At order $1/x^2$ the replica calculation introduces only one new combination
%which is uniquely determined by the result for $\nu =1/2$. At order $1/x^3$,
%two new combination enter. They are uniquely determined by the result
%for $\nu=1/2$ and $\nu=3/2$. At order $1/x^4$ four new combination 
%enter in the replica calculation. They cannot be determined from
%the results for  $\nu=1/2$, $\nu=3/2$
%and $\nu = 5/2$.

\section{Spectral Density via Fermionic Replicas}

In this section, we analyze    the low-energy partition
in the quenched case ($N_f =0$). 
~For $n$ replica flavors in the sector
of topological charge $\nu$ it is given by (see (\ref{zcomp})) \cite{LS}   
\be
Z_{\nu}^{(n)}(x) = \int_{U\in U(n)}
 dU {\det}^\nu U e^{\frac {x}{2} {\rm Tr} (U + U^{-1})}.
\label{z2}
\ee
In order to take the replica limit, $n\to 0$, the $n$-dependence of  
(\ref{z2}) has to be known explicitly.
The closed form of  $Z_{\nu}^{(n)}(x)$, in terms of a determinant of modified 
Bessel functions,
\be
Z_{\nu}^{(n)}(x)= \det(I_{\nu +j-i}(x)),\,\, i,j=1,\cdots , n,
\label{zls}
\ee
does not provide us with an explicit $n$-dependence. 
The explicit $n$-dependence
can only be obtained for the large-mass and the small-mass expansion
of $Z_{\nu}^{(n)}(x)$. It was obtained in \cite{kim,poul} for an
expansion about the replica symmetric saddle point using the method
of Virasoro constraints.  Our new result is 
the asymptotic expansion of the microscopic spectral density 
using replica symmetry
breaking \` a la Kamenev and M\'ezard \cite{km}
which will be discussed the the second
half of the next subsection. 
The small mass expansion of (\ref{zls}) was also considered in 
\cite{kim,poul},
but failed at the order for which logarithmic terms enter in the expansion.
In the second subsection we derive the lowest order
logarithmic term  for the case of zero topological 
charge. 

\subsection{Large Mass Expansion}

The asymptotic expansion of the microscopic spectral density is
obtained from the large mass expansion of the finite volume 
partition function. To this end we expand the partition function
(\ref{z2}) in powers of $1/x$ by means of a saddle point 
approximation. By diagonalizing $U$ it can be easily seen that
the saddle points are given by unitary matrices with eigenvalues
$\pm 1$, i.e. by unitary matrices satisfying $U^2 =1$. 
The solutions of this equation 
are highly degenerate. They can be organized in $n+1$ 
classes,  $U=I_p$,   where $I_p$ is a diagonal
matrix with $p$ elements  $-1$ and $n-p$ elements $+1$ (with $0\le p \le n$).
The integrand does not depend on the submanifold
 $U(n)/U(n-p)\times U(p)$ of $U(n)$ and the 
integration over this coset has to be performed exactly
resulting in its volume $V_{n,p}$. In terms of the parameterization
\be
U = I_pU_0 V U_0^{-1},
\ee
with $U_0 \in U(n)/U(n-p)\times U(p)$ it is clear that the 
integration can be restricted to $U(n-p)\times U(p)$. 
%For each integer $p$ there are  
%$\left ( \begin{array}{c} n \\ p 
%\end{array} \right )$ degenerate saddle points. 
Summing over all saddle points the partition function is given by
\be
Z_{\nu}^{(n)}(x) =  \sum_{p=0}^{n}
V_{n,p}
 {\int_{V\in U(n-p)\times U(p)} }
 dV J(V) {\det}^\nu V e^{\frac {x}{2} {\rm Tr}[I_p (V + V^{-1})]}\label{zsum}.
\ee
where $V_{n,0}=1$ and for $p\ne 0$ 
\be 
V_{n,p}=(2\pi)^{p(n-p)}
 \left ( \begin{array}{c} n \\ p \end{array} \right)
\frac{ \prod_{\j=1}^p j!\prod_{\j=1}^{n-p} j!}
{\prod_{\j=1}^n j!}\,\equiv \, (2\pi)^{p(n-p)}F_n^p\label{fnp}.
\ee
The integration over $V$ 
should be thought of  a saddle-point integral 
of a formal expansion of $V$ about the identity to all orders. 
Below we will make this explicit for the different types of saddlepoints.
The total number of saddle-points in the class $p$ is $\frac {n!}{(n-p)!p!}$.
This factor is included as combinatorial factor in $V_{n,p}$. The volume of the
coset $U(n)/U(n-p)\times U(p)$
is given by the ratio  $\frac {V(U(n))}{V(U(p))\, V(U(n-p))}$. With the
volume of $U(k)$ given by   $(2\pi)^{\frac {k(k+1)}{2}}\prod_{j=1}^k j!$
we obtain the volume factor $V_{n,p}$.

~From the exact expression of the partition function (\ref{z2}) 
in terms of modified
Bessel function given in (\ref{zls}) it is clear that the asymptotic
series of $Z^{\nu}_n(x)$ terminates for half integer $\nu$. 
Let us investigate the asymptotic expansion about the saddle point $I_p$
in more detail. Because the total number of 
degrees of freedom in $U(n-p)\times U(p)$ is equal to $(n-p)^2 + p^2$,
the expansion of $Z_{\nu}^{(n)}(x)$ is of  the form
\be
Z_{\nu}^{(n)}(x) \sim e^{(n-2p)x}
\left ( \frac 1x \right )^{(n-p)^2/2 + p^2/2}
(1+{\cal O} \left(\frac 1x \right))\label{desperate}.
\ee
This result is valid for arbitrary $\nu$. The asymptotic series of
the Bessel function $I_{k+\frac 12}(x)$ terminates at $1/x^{k+1/2}$. 
The result for   the
maximum power in $1/x$ occuring the expansion
of  the determinant (\ref{zls}) is particularly simple for $\nu = \frac12$
and is given by
\be
\left ( \frac 1x \right )^{n^2/2 }.
\label{only}
\ee
Let us  consider this case  
in more detail. Since,
as we will see below, only the saddle points for $p=0$ or $p=1$ 
contribute in the replica limit, we only discuss these values of $p$. 
We observe that the maximum power and the minimum power in the asymptotic
series are equal for $p=0$ saddle point. 
The asymptotic series thus has only one term and we find that
\be
\left(Z_{\nu=\frac 12}^{(n)}(x)\right )_{p=0} 
&\sim& e^{nx}\left ( \frac 1x \right )^{n^2/2}.
\label{heck}
\ee
~For $p=1$ and 
larger half-integer values of $\nu$, more terms contribute to the
asymptotic series, but it still terminates. 
~For example, for $p=1$ and $\nu =\frac 12$ one finds from (\ref{desperate})
and (\ref{only}) that the difference between the maximum and minimum 
power in the expansion in $1/\sqrt x$ is $n-1$. Therefore, the  
asymptotic series for $Z^{\nu}_n(x)$ 
cannot contain more than $n$ terms.
~From numerical examples for small values of $n$, one indeed finds
that in this case the coefficients of all  
$n$ possible terms are nonvanishing. 
~For the same reason as in the case of the GUE two-point
function \cite{zirn}, we expect that the replica trick will give the
exact result both for $\nu =\frac 12$ as well as larger 
half-integer values of $\nu$.

Clearly, the expression (\ref{zsum}) makes only sense for 
positive integer values of $n$ . In order
to analytically continue it we follow the work of
Kamenev and M\'ezard \cite{km}.
They analytically continued the factorials in $F_n^p$ such that
 $F_n^p$ vanishes  for $p\ge n+1 $. Then the sum in (\ref{zsum}) can be
extended up to infinity and the replica limit $n\to 0$ 
can be taken term  by term in the sum over $p$. One finds
\be
\lim_{n\rightarrow 0} F_n^p  \sim n^p,
\ee
so that only the terms $p=0$ and $p= 1$ 
of the,  to infinity continued,  sum in (\ref{zsum})
survive. In these two cases  we obtain ${ F}_{n\to 0}^0=1
\, , \,{ F}_{n\to 0}^1=n $.  
 Notice, that 
 the continuation of the sum over $p$ to infinity 
explicitly breaks the replica symmetry $p \rightarrow n-p$.
Of course, the group integral in (30) must be also 
continued to non-integer values of $n$. 

The previous discussion suggests the definition 
\be 
Z_{\nu}^{(n)}(x) \, \equiv \, \left(Z_{\nu}^{(n)}(x)\right)_{p=0}
\, + \, n\, \left(Z_{\nu}^{(n)}(x)\right)_{p=1}.  
\ee
We will first consider the contribution for $p=0$ which originates from
expanding $U$ around the identity matrix $I_0$.
Although not necessary, it turned out 
to be convenient to parameterize $V_{p=0}\in U(n)$ according to
\be
V_{p=0} = \frac{1+iH/2}{1-iH/2},
\ee
where $H$ is an Hermitian $n\times n$ matrix. ~From a diagonal representation
of $V_{p=0}$ one can easily show that \cite{pietbrouwer}
\be
d V_{p=0} = \frac 1{{\det}^n(1+H^2/4)} dH.
\label{jac0}
\ee
In the replica limit,  the Jacobian can  be ignored and one simply has
$d V_{p=0} =  dH$. The $p=0$ contribution to the partition function
is thus given by
\be
\left(Z_{\nu}^{(n)}(x)\right)_{p=0} = 
\int &dH&  \,{\det}\left ( \frac {1+iH/2}{1-iH/2} \right )^{\nu}  
 e^{n x -\frac x2{\rm Tr}H^2  
+ x{\rm Tr} \left(\frac{H^4/8}{1+H^2/4}\right )}.  
\ee
In order to keep track 
of the powers of $x$ it is convenient 
to rescale $H\to H/\sqrt{x} $ which leads to another
Jacobian $x^{-n^2/2}$ that also vanishes in the replica limit.
In terms of $x=m\Sigma_0 V$, the mass dependence of the condensate is
given by 
\be
\frac{\Sigma_{p=0}(x)}{\Sigma_0}
\, =\,  \lim_{n\to 0}
\frac{1}{n}\frac 1{Z_{\nu}^{(n)}(x)}
\frac{\partial}{\partial x}\left(Z_{\nu}^{(n)}
(x)\right)_{p=0}.
\label{res}
\ee
so that to order $1/x^4$ we need to collect terms to order $1/x^3$
in the expansion of the partition function. Using the
expressions
\be
e^{\frac {x}2  Tr(V_{p=0} + V_{p=0}^{-1})} &=& e^{nx -\frac 12
{\rm Tr}H^2}\, e^{{\rm Tr}
\left\lbrack \frac {1}{8x} H^4 -\frac {1}{32x^2}H^6 
+\frac {1}{128x^3} H^8 -
\cdots \right\rbrack }\nonumber\\
(\det V_{p=0})^{\nu}\, &=&\, e^{\nu n}e^{\frac{i\nu\,{\rm Tr}}
{\sqrt{x}}
\left(H - \frac{2H^3}{3\times 2^3 x^2}+ 
\cdots \right)}\label{expan}.
\ee 
we find the result (only terms up to order $1/x^2$ are displayed)
\be
\left(Z_{\nu}^{(n)}\right)_{p=0}&=& \frac {e^{nx+\nu n}}{x^{\frac {n^2}{2}}}
\int dH e^{-\frac 12 {\rm Tr} H^2}\nonumber \\
&\times&(1 - \frac {\nu^2}{2x} ({\rm Tr}H)^2
+\frac {\nu^2}{12x^2} ({\rm Tr}H)({\rm Tr}H)^2
+\frac {\nu^4}{24 x^2} ({\rm Tr}H)^4 ) \nonumber \\ &\times &
(1+ \frac 1{8x} {\rm Tr}H^4-\frac 1{32x^2} {\rm Tr}H^6)\nonumber \\
&=& \left( \frac{2\pi}{x}\right)^{n^2/2} 
e^{n x}( 1 - n \frac{\nu^2}{2 x} +
(2n^3+n) \frac 1{8x})\label{zp0} + {\cal O} (\frac{n^2}{x^2}).
\ee
The Gaussian integrals have been calculated using the 
trace correlators given in the
Appendix. To this order, the Jacobian in (\ref{jac0}) gives rise to an
extra factor $(1-n^3x/4)$. With inclusion of this terms the $1/x$
corrections vanish for $\nu = \frac 12$. 
All terms of order $1/x^2$ are at least of order $n^2$.
The resolvent  obtained from (\ref{res})
\be 
\frac{\Sigma_{p=0}(x)}{\Sigma_0}
\, =\, 1\, + \, \frac {4\nu^2-1}{8x^2} \, - \,
\frac {(4\nu^2-1)(4\nu^2-9)}{128x^4} \, + \, \cdots
\label{notosci}
\ee 
 agrees with the large mass expansion
of the quenched condensate obtained via other methods \cite{kim}.
We observe a cancellation of the odd powers in $1/x$. This is in
agreement with the analytical expression for $\Sigma(x)$ 
given in (\ref{valsup}) which
can be rewritten as
\be
\frac {\Sigma(x)}{\Sigma_0}
= \frac x2 ((2-\frac{4\nu^2}{x^2})I_\nu(x)K_\nu(x) + I_{\nu+1}(x)K_{\nu+1}(x)
+ I_{\nu-1}(x)K_{\nu-1}(x)).
\ee
~From the asymptotic behavior of the Bessel functions, one easily derives
that the asymptotic series of this expression
is an expansion in powers of $1/x^2$.
Also notice the cancellation of the $\nu/x$ term in the large$-x$
asymptotic expansion.
 
Next we consider the more subtle 
contribution of the saddle point given by the diagonal
matrix $I_1$ with one element equal to $-1$ and $n-1$ elements equal to $1$.
A parameterization of $U(n-1) \times U(1)$ that is convenient 
for the expansion about the saddle point is given by 
\be
V_{p=1} = \left ( \begin{array}{cc} \frac{1+iH/2}{1-iH/2} &   \\  &
              \frac{1+ih/2}{1-ih/2} \end{array} \right ),
\ee
where, now, $H$ is a hermitian 
$(n-1)\times (n-1)$ matrix 
and $h$ is a real variable. Because $U_0 \in U(n) /U(n-1)\times U(1)$
we have that
\be
U = I_1U_0 V_{p=1} U_0^{-1}= U_0 I_1 V_{p=1} U_0^{-1}.
\ee
The measure, in terms of the coordinates $H$ and $h$, can be obtained
by diagonalizing $U$ and $H$ with unitary transformations $U_1$ and $W$,
respectively. If the eigenvalues are denoted by $e^{i\theta}$ 
and $h_k$, in this
order, we find the measure (notice the plus sign in the last factor)
\be
dU &=& \prod_{k< l} |e^{i\theta_k} - e^{i\theta_l}|^2 \prod_k d\theta_k dU_1
\nonumber \\ &=&
     \frac{\prod_{k<l}|h_k-h_l|^2}{(1+h^2/4)\prod_k(1+h^2_k/4)^{n-1}}
     \prod_k\left  
| \frac{1+ih_k/2}{1-ih_k/2} + \frac{1+ih/2}{1-ih/2}\right |^2 dh \prod_k dh_k 
\,dU_0 d V. \nonumber \\
\ee
The last factor can be written as
\be
\prod_k\left  | \frac{1+ih_k/2}{1-ih_k/2} + \frac{1+ih/2}{1-ih/2}\right |^2
= 4^{n-1}\frac{ {\det}^2(1+ h H)}{(1+h^2/4)^{n-1} \det(1+H^2/4)}.
\ee
Using that $\prod_{k<l}|h_k-h_l|^2\prod_k dh_k\, dV = dH$ we thus find
the measure
\be
dU =  
4^{n-1}dU_0\frac { dh dH  {\det}^2(1+ h H)}{(1+h^2/4)^{n} {\det}^n(1+H^2/4)}
\equiv J(H,h) dh dH dU_0. 
\label{jacob}
\ee
The integrand does not depend on $U_0$ and the integration over these variables
just gives the volume of the coset  which, together with the combinatorial
factor, combines into the factor $V_{n,1}$ discussed in the first part of this
subsection. The
 $p=1$ contribution  to the partition function is thus given by 
 (with one overall minus sign from orienting the $h$-integration from
$-\infty$ to $\infty$) 
\be
\left(Z_{\nu}^{(n)}(x)\right)_{p=1} &=& 
-(-1)^{\nu} (8\pi)^{n-1}\int dH dh \, 
\frac{{\det}^2(1+hH)}{(1+h^2/4)^n{\det}^{n}(1+H^2/4)}
\nonumber \\&&\times
 \,\left ( \frac{1+ih/2}
{1-ih/2} \right )^\nu 
{\det}\left ( \frac {1+iH/2}{1-iH/2} \right )^{\nu}  
  e^{x{\rm Tr} \left(\frac{1-H^2/4}{1+H^2/4}\right ) 
-x\left(\frac {1-h^2/4}{1+h^2/4}\right)}.  
\ee
Since the factors in the denominator of the Jacobian can be ignored in the
replica limit only the following terms in the expansion of the 
Jacobian contribute to order $1/x^4$,
\be
J=
4^{n-1}\left\lbrack 1 + \frac{h{\rm Tr}H}{2} + \frac{h^2}{16}\left\lbrack
2({\rm Tr}H)^2 - {\rm Tr}H^2\right\rbrack + \cdots \right\rbrack.
\ee 

It is instructive to perform the calculation 
to leading order in $1/x$. In this case 
the following terms should be collected 
\be
\left(Z_{\nu}^{(n)}(x)\right)_{p=1} &=& 
-(-1)^{\nu } (8\pi)^{n-1}e^{(n-2)x+\nu n}\int dH dh
e^{-\frac {x}2 {\rm Tr} H^2 +\frac {x }2 h^2} 
\nonumber \\ &\times&
\left [ 1 + \frac {x}8 {\rm Tr} H^4
- \frac {x}8 h^4 
-\frac{\nu^2}2 (({\rm Tr}H)^2 + h^2 + 2ih{\rm Tr}H )
+ \frac{h{\rm Tr}H}{2}   \right ].\label{zp1} 
\nonumber \\
\ee
The last two terms vanish upon integration. Thus, the Jacobian 
contributes only at the next to the leading order.
The saddle-point integrations can be performed conveniently
by rescaling $h$ and $H$
according to  $h\to h/\sqrt{-x}\, ; \, H\to H/\sqrt{x}$.  
This results in
\be
\left(Z_{\nu}^{(n)}(x)\right)_{p=1} &=& 
i (-1)^{\nu}(8\pi)^{n-1}
\left ( \frac{2\pi}x \right )^{((n-1)^2 +1)/2} 
 e^{(n-2)x+\nu n}\non\\
&\times& \left [1+ (2(n-1)^3+(n-1))\frac 1{8x}-\frac 3{8x}
-n(n-2)\frac{\nu^2}{2x} \right].
\ee
Now we are in a position to calculate the contribution of the 
$p=1$ saddle point to the chiral condensate,
\be
\frac{\Sigma_{p=1}(x)}{\Sigma_0}
\, =\,  \lim_{n\to 0}
\frac{1}{n}\frac 1{Z_{\nu}^{n}(x)}
\frac{\partial}{\partial x}\lbrack n\, \left(Z_{\nu}^{(n)}(x)
\right)_{p=1}\rbrack.
\ee
Using trace correlators given in the Appendix, the expansion in $1/x$ can
be easily extended to order $1/x^4$. ~For the mass dependence of the
chiral condensate we obtain
\be
\frac{\Sigma(x)}{\Sigma_0}
\, &=&\,  \lim_{n\to 0}
\frac{1}{n}\frac 1{Z_{\nu}^{n}}(x)
\frac{\partial}{\partial x}Z_{\nu}^n(x)\nonumber\\
&=& 1 -\frac{i(-1)^{\nu}e^{-2x}}{2x} + 
\frac{(4\nu^2-1)(1-i(-1)^{\nu}e^{-2 x})}{8x^2}
-\frac{i(-1)^{\nu}e^{-2x}(4\nu^2-1)(4\nu^2-9)}
{64x^3}\nonumber\\ 
&+& \frac{(4\nu^2-1)(4\nu^2-9)\left\lbrack
i(-1)^{\nu}e^{-2x}(19-4\nu^2)-6\right\rbrack}{3\times 256 x^4}.
\label{cond4}
\ee
All  terms $\sim i e^{-2x}$ originate from $p=1$ saddle point.
Before calculating the spectral density by  taking the
discontinuity of $\Sigma(x)$ we wish to point out that
the expression (\ref{cond4}) has been obtained under the assumption
 that ${\rm Re}( x)>0$, so that we cannot calculate the 
discontinuity from the difference of $\Sigma(i\lambda +\epsilon)$ and
$\Sigma(i\lambda -\epsilon)$. The reason is that for  $x = i\lambda +\epsilon$
the dominant saddle-point is given by $I_0$ and for $x = i\lambda -\epsilon$
it is given by $-I_0$.
The infinitesimal increment thus breaks
the replica symmetry between $I_0$ and $-I_0$.
The sum over $p$ in (\ref{zsum}) has been extended to $\infty$ consistent
with the breaking of the replica symmetry $p\to n-p $ by the saddle point
$I_0$.
Thus, the infinitesimal increment is necessary 
to resolve the ambiguity as was also the case in the original 
calculation of \cite{km}. The expression for 
$\Sigma (i\lambda -\epsilon)$ can be obtained
from a replica-symmetry breaking solution for which $-I_0$ dominates.
%Alternatively, we can calculate the discontinuity according to
%(\ref{discrho}) without the need of a replica dual calculation.
The final result for for the asymptotic
expansion of $\rho_s(\lambda)$  coincides
with (\ref{rho4}) obtained from
the expansion of the analytical result \cite{VZm}
$(\lambda /2)(J_{\nu}^2(\lambda)-J_{\nu+1}(\lambda)
J_{\nu -1}(\lambda))$ up to $1/\lambda^4$ for $\lambda>0$.

%\be
%\rho_\nu(x) &=&
%\frac 1{\pi }\left\lbrack 1 - \frac{ \cos (2x-\pi\nu)}{2x} + 
%\frac{1-4\nu^2}{8x^2} (1-\sin (2x -\pi\nu))\right .
%\nonumber \\&+& 
%(4\nu^2-1)(4\nu^2-9)\frac{\cos (2x-\pi\nu)}
%             {\pi 64 x^3}\nonumber \\
%&+& \left . (4\nu^2-1)(4\nu^2-9)\frac{( -6 + (19-4\nu^2) \sin (2x-\pi\nu )}
%                    {\pi x^4 2^7 3!} + \cdots \right\rbrack .
%\label{rho4'}
%\ee
We observe that it requires a great deal of effort to derive the asymptotic
expansion of the oscillating contribution to the spectral density by means
of the replica trick. This is
especially true due to the lack
of the Virasoro constraints for 
$\nu\ne 0 $. Those constraints are
a key tool \cite{kim} in simplifying the calculations 
for the mass expansions of the 
condensate in the sector of vanishing
topological charge ($\nu=0$). 
The simplifying  feature in the sector 
with vanishing topological charge  that 
the partition function can be shown to 
belong to the  universality class (see \cite{mironov}) 
of the generalized Kontsevich model with potential 
${\cal V}(x)=1/x^2 $ and satisfies the same 
Virsoro constraints. ~For $\nu\ne 0$
the one-link integral depends also
on $\det J $ and $\det J^{\dagger }$
which is also the case when we have 
integrals over $SU(N_f)$ instead of $U(N_f)$. 
Therefore, a possible generalization  of the Virasoro constraints
to include the case $\nu\ne 0$ would be certainly 
welcome not only to  reduce
our efforts in perturbative calculations 
but also as an identification of the 
universality class of the QCD finite volume partition function at
nonzero topological charge.

\subsection{Small Mass Expansion}

The situation for the small mass is much more complicated.
Before discussing the complications we
recall that the partition function $Z_{\nu}^n(x)$
has been extensively studied in the context
of lattice QCD where it is known as the one-link
integral (see for example \cite{rossi} for an updated review
of the subject). In that context one considers unitary matrix integrals
of the form
\be
Z(J,J^\dagger) = \int_{U\in U(n)} dU e^{
 {\rm Tr }\left( J U^{\dagger} + J^{\dagger}U\right)},
\ee
where  $J$  is a general $n\times n$ 
matrix. The partition function  with such potential 
is  a function of the eigenvalues 
$\lambda_k$  of the matrix  $JJ^{\dagger}$. 
The one-link model  exhibits two phases according to \cite{GW} 
\be
\sum_{k=1}^n \frac{1}{2\sqrt{\lambda_k}}\, &&\, 
  \le  1 \quad ({\rm weak \,\, coupling}),\non\\
  \sum_{k=1}^n \frac{1}{2\sqrt{\lambda_k}}\, 
&& \ge  1 \quad ({\rm strong \,\, coupling}).\non\\
\ee
In our  case, with partition function given by (\ref{zls}), the eigenvalues
of $J^\dagger J$ are given by $x^2/4$.
~From  the above we see that, for $x\to\infty$,
we expect to take the replica limit, $n\to 0$, and remain
in the weak coupling regime while, for $x \to 0$,  it is not clear 
whether the replica limit can be taken without crossing a phase boundary.
These problems are reflected in logarithmic singularities of
the small $x$ expansion of the
valence quark mass dependence of the partition function and its
derivatives. In \cite{kim} the replica limit
of the expansion coefficients could be derived up to the order
for which terms of the form $x^p\log x$ are absent. These singular
terms could be related \cite{kim} to the presence of de~Wit-'t Hooft
poles \cite{hooft}.

Below we consider the replica limit of $Z_\nu^n(x)$ for $\nu = 0$. In that
case logarithmic terms already enter to lowest order in the expansion.
We thus consider the small $x$ expansion of the partition function
$Z_{\nu =0}^n(x) = \det (I_{i-j}(x))$. By expanding the Bessel functions
we obtain
\be
Z_{0}^n (x)\, 
=\, \sum_{k=0}^n \left(\frac{x^2}{4}\right)^k\frac{1}{k!} \, +\,
\sum_{k=n+1}^\infty C_{k,n}x^{2k} .\ee
The first sum can be recognized as an incomplete exponential, but
we were not able to determine the coefficients $C_{k,n}$ in general.
In order to expose the $n$-dependence of the first term, we rewrite
the incomplete exponential as an incomplete gamma function 
\be
\sum_{k=0}^n \left(\frac{x^2}{4}\right)^k\frac{1}{k!}
\, = \, \frac{e^{\frac{x^2}{4}}\Gamma(n+1,x^2/4)}{\Gamma(n+1)}.
\ee
~From the small mass expansion  of the incomplete
gamma function 
\be
\Gamma(n+1,x^2/4) \, = \,\int_{x^2/4}^{\infty}dt \,t^n e^{-t} \, =\,
\Gamma(n+1) \, + \, \sum_{k=1}^{\infty} 
\left(\frac{x^2}{4}\right)^{k+2n}\frac{(-1)^k}{(n+k)(k-1)!}, 
\ee
and neglecting terms which will vanish
in the replica limit, i.e., 
\be
\Gamma (n+1)&=& 1- \gamma n \, + \, {\cal O}(n^2)\non\\
\left(\frac{\mu^2}{4}\right)^n\frac{1}{n+k}\, &=&\,
\frac{1}{k} \, + \, n\left( \frac{2}{k}\log\frac{\mu}{2}
-\frac{1}{k^2} \right) + {\cal O}(n^2), 
\ee
we find
\be
Z_0^n(x) \, &=& \, 1 \, + \, n\left\lbrack (1-e^{x^2/4})(2\log\frac{x}{2}
+\gamma)  + e^{x^2/4}\sum_{k=1}^{\infty}
 \left(-\frac{x^2}{4}\right)^k\frac{1}{k!k} \right\rbrack
+{\cal O}(n^2) \non\\
&+& \,\, \sum_{k=n+1}^{\infty}C_{k,n}x^{2k}. \non\\
\ee
Because the coefficients $C_{k,n}$ are unknown only the
lowest order terms of the small mass expansion can be calculated,
\be
Z_0^n(x) \, &=& \, 1 \, - \, n \frac {x^2}2 \log x
+ {\cal O}(x^2) + {\cal O}(n^2). 
\ee
~For the replica limit of the condensate  given by,
\be
\Sigma (x ) \, = \, 
\lim_{n\to 0}\frac{1}{nZ_0^n(x)}\frac{\del Z_0^n(x)}{\del x},
\ee 
we thus obtain
\be
\Sigma (x) \, =\, - x \, \log x \, + \, {\cal O}(x^1) .
\ee
This is indeed the correct leading order term of the 
small mass expansion of $x(I_0(x)K_0(x) + I_1(x)K_1(x))$.
The linear behavior of the microscopic spectral density at the origin 
is reproduced by taking the discontinuity 
across the imaginary axis,
\be
\rho_s(\lambda)=\frac{\lambda}{2} + {\cal O}(\lambda^2).
\ee

We did not succeed to generalize this calculation to arbitrary $\nu$,
but we expect that the logarithmic terms can be obtained in a 
similar fashion. In particular, the first $n+\nu$ terms of the expansion
seem to follow a simpler pattern than the coefficients of the higher
powers.

\section{Bosonic Replicas }
In view of the seemingly different role
\cite{VZ,zirn} played by bosonic and fermionic replicas 
and the fact that the supersymmetric method uses both
compact and  noncompact variables it is natural to try 
to reproduce the results of last section by introducing
additional $n$ replicas of bosonic quarks of mass
$m$ instead of fermionic ones. In this case, the
condensate for $N_f=0$ is given by 
\be 
\Sigma(m)
\, =\,  \lim_{n\to 0}
\frac{1}{(-n)}\frac 1{V_4} \frac{\partial}{\partial m}
\ln { Z}_{\nu}^{(-n)}(x) ~,\label{condbos}
\ee
where ${ Z}_{\nu}^{(-n)}(x)$ is given in (\ref{z1}). 
We will show that, for large masses,
bosonic replicas can be used to reproduce the asymptotic expansion of the
chiral condensate but they fail to reproduce the microscopic
spectral density for a subtle reason which we will explain below.

It is convenient to express the bosonic partition function 
$Z_{\nu}^{(-n)}(x)$
in terms of the eigenvalues of the $Gl(n)/U(n)$ matrices.
Up to an overall constant we have,
\be
Z_{\nu}^{(-n)}(x) = \int_{-\infty}^{\infty}
\prod_{k=1}^n d s_k \prod_{k<j}(e^{s_k} - e^{s_j})
(e^{-s_k} - e^{-s_j})e^{-x\sum_
{k=1}^n \cosh s_k + \nu \sum_{k=1}^n s_k }. \label{z-n}
\ee
Notice in particular that the measure (including 
the Vandermonde determinant) is invariant under the 
symmetry $s_k\to s_k +t $ which is 
a remnant of the $Gl(n)/U(n)$ invariance. 
The fermionic partition function $Z_{\nu}^{(n)}(x) $ is given
by a circular ensemble (for integer $\nu$) \cite{LS},
\be
Z_{\nu}^{(n)}(x) = \int_{-\pi}^{\pi}
\prod_{k=1}^n d \theta_k \prod_{k<j}(e^{i\theta_k} - e^{i\theta_j})
(e^{-i\theta_k} - e^{-i\theta_j})e^{x\sum_
{k=1}^n \cos \theta_k + i\nu \sum_{k=1}^n \theta_k }. \label{z-c}
\ee

In the noncompact case, the solutions of  the saddle point equation
$\sinh s_k = 0 $ are given by $s_k = 0,\pm i\pi ,\pm i2\pi
\cdots $. Thus, in principle we might have a variety
of saddle points which should be all taken into account
in  large $x$ expansion. However, we will argue that only $s_k=0$ solution
contributes to the large $x$ behavior of $Z_{\nu}^{(-n)}(x)$.
Our discussion is based on the $n=1$ integral where a 
steepest descent analysis can be easily carried out. 
In this case, the bosonic partition function is given by
\be
Z_{\nu}^{(-1)}(x) =  
\int_{-\infty}^{\infty}ds e^{-x\cosh (s)+\nu s}= 2 K_\nu(x) 
\ee
~From the asymptotic expansion of $K_\nu(x)$ it is clear that only the 
saddle-point at $s =0$ contributes to the integral. 
The spectral density  can be calculated
from the resolvent at  $x=\pm i\lambda +\epsilon $ so that the 
integral above can be identified with the modified Bessel 
function $ K_\nu(x)$.

\begin{center}
\begin{figure}[ht!]
\begin{center}
\hspace*{0.3in} 
{\epsfig{figure=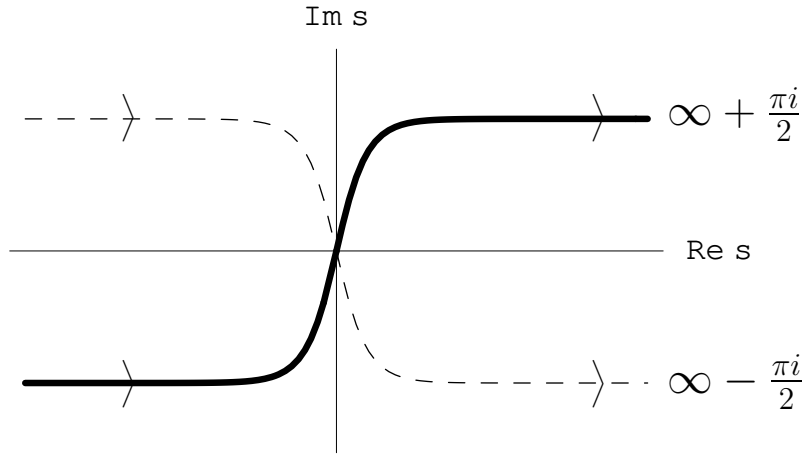,width=100mm}
\setlength{\unitlength}{2.0in} 
\begin{picture}(1,1)                                                  
%\put(-0.3,0.87){\bf\Large $\infty +\frac {\pi i} 2$}
%\put(-0.3,0.17){\bf\Large $\infty-\frac {\pi i} 2$}
\put(1.33,1.88){\bf\Large $\infty +\frac {\pi i} 2$}
\put(1.33,1.18){\bf\Large $\infty-\frac {\pi i} 2$}
% \put(-0.65, 0.87){${\bf\Big >}$}
% \put(-0.65, 0.175){${\Big >}$}
% \put(-1.85, 0.87){${\Big >}$}
% \put(-1.85, 0.175){${\bf\Big >}$}
 \put(1.13, 1.88){${\bf\Big >}$}
 \put(1.13, 1.183){${\Big >}$}
 \put(-0.1, 1.88){${\Big >}$}
 \put(-0.1, 1.183){${\bf\Big >}$}
\end{picture} }
\end{center}     
\vspace*{-4cm}                                      
\caption{Steepest descent curve passing through the saddle point $s=0$
for $\lambda < 0 $ (solid) and for $\lambda >0$ (dashed),
where we assumed $x=i\lambda +\epsilon$.}
\label{steep}
\end{figure}
\end{center}

On the other hand, for one fermionic replica the partition function
is given by (assuming integer $\nu$)
\be
Z_{\nu}^{(1)}(x) = 2\pi I_\nu(x).
\ee
In this case two saddle-points contribute to the asymptotic expansion, one
$\sim e^x$, and the other one $\sim e^{-x}$, which are both essential to 
recover the oscillatory contributions to the 
microscopic spectral density.

Let us analyze in detail the saddle-point calculation of the
bosonic partition function for one replica. The stationary phase condition
that the imaginary part of the action, $-i\lambda \cosh s$, 
is constant results in  the following curve in the complex $s$-plane 
through the saddle point at $s = 0$.
Clearly, the integration contour 
(the real axis) can be deformed into the steepest
descent curve of fig. 2 (depending on the sign of $\lambda$)
and no other saddle points need to be considered.
A similar analysis for the compact case shows that both the saddle
points at $\theta =0$ and at $\theta = \pi$ have to be taken into account.

A similar situation arises in the $1/N$ correction to the semicircle law
of the Gaussian Unitary Ensemble (GUE) of hermitian matrices.
Inside the semicircle the saddle point solutions of the $n=1$ fermionic
replica are horizontally aligned in the complex plane, 
while  (see also a comment in \cite{zirn})
 the solutions for the one bosonic replica are
vertically aligned and only the $p=0$ saddle point contributes.
If we insist on taking into account all vertically 
aligned saddle points we get an incorrect result. However,
we have checked that outside
the semicircle the situation is reversed (see figures in \cite{Cava}
for the fermionic case). In this case 
the bosonic replicas correctly reproduce the 
semicircle law and its leading exponentially decreasing 
 correction outside the semicircle (see also \cite{parisi98})
while the fermionic replica calculation gives
an incorrect result. In the case of the microscopic
spectral density analyzed in this work,
we are always inside the semicircle even
in the limit $x\to \infty$ which explains, assuming the same pattern of 
as for the GUE, 
the failure of the bosonic replica calculation 
in reproducing the oscillating part of 
the spectral density.

If we are just interested in the mass dependence of the
chiral condensate it is 
sufficient to only take into account the $p=0$ saddle point and, one can 
convince oneself that bosonic and 
fermionic replicas produce the same large 
$x$ result as follows.
The integrand of the non-compact partition function is obtained
by transforming the integration variables according to $\theta_k = is_k$
and replacing $x \to -x$.
~For large $x$, the saddle-point of both partition functions is at
$\theta_k =0$ and $s_k = 0$, respectively, and one finds 
that $Z_{\nu}^{(n)}(-x)$ and $Z_{\nu}^{(-n)}(x)$ have  
the same asymptotic
expansion  for the chiral condensate (see  (\ref{notosci})).
Clearly this proof is formal because it assumes that the replica
limit can be interchanged with the operation $x\to -x$.
However, we have  explicitly checked to the order $1/x^3$ 
that the noncompact partition function $Z_{\nu}^{(-n)}(x)$
leads to the same asymptotic expansion. 

\section{Conclusions}

We have investigated the replica trick for the microscopic spectral
density of the QCD Dirac operator in the quenched limit. The advantage
of working with fermionic replicas is that this theory corresponds to
QCD with $n$
flavors of equal mass $m$. Because the low energy properties of this theory
have are well  understood, the starting point of this approach 
has a firm basis. The valence quark mass dependence of the chiral
condensate and the spectral density of the QCD Dirac operator, however,
are only obtained in the limit $n \to 0$. The existence of this limit
has been debated for many years and the investigation of its nature
has been the main topic of this article.

The alternative approach to obtain the QCD Dirac spectrum is the 
supersymmetric method. Although in principle rigorous, one might
raise the question whether this theory with bosonic ghost quarks might
have unusual properties as for example the spontaneous breaking of 
supersymmetry. Our results show that this is not the case. The 
low-energy limit of this partition function is completely dictated
by chiral supersymmetry. The power of the supersymmetric method
is that one obtains rigorous nonperturbative results
such as, for example, the spectral density in the microscopic region. 

The replica trick, on the other hand, requires an explicit $n$ dependence,
and, up to now, only perturbative results have been obtained.
Exact results have only been derived in cases where the
perturbative series consists of only a finite number of terms.
Our results for the microscopic spectral density confirm that the asymptotic
series of its non-oscillatory part 
can be obtained from an expansion about the replica symmetric
saddle-point. Both fermionic and bosonic replicas give the same result
and are in complete agreement with the asymptotic expansion of the exact 
result. 
The same is true for the asymptotic expansion of the
resolvent away from the imaginary axis (where the eigenvalues are located).
Things are different for the the oscillatory part of the 
microscopic spectral density. In this case the correct asymptotic
expansion is obtained only if a saddle point that breaks
the replica symmetry is taken into account. This additional
saddle point only exists for fermionic replicas. ~For bosonic replicas,
only the replica symmetric saddle point contributes in the saddle
point calculation, and therefore this approach does not 
reproduce the asymptotic
expansion of the oscillatory part of the spectral density. A similar
observation has been made for the application of the replica trick to
the Wigner-Dyson ensembles.
   
In the asymptotic expansion of the supersymmetric partition function
also two saddle-points have to be taken into account. 
In the boson-boson component of the Goldstone manifold only one
saddle point contributes, but
as is the case for one fermionic replica, we have to take into
account two saddle-points in the fermion-fermion component of
the Goldstone manifold.

Is it possible to go beyond perturbation theory using the replica
trick? One indication in favor of an affirmative answer to this
question is that we have reproduced the leading order logarithmic singularity
of the small mass expansion of the resolvent. Except for these 
logarithmic terms, the small mass expansion of the resolvent is a convergent
series which may be summed to obtain the exact result.
The large mass expansion, on the other hand, is an asymptotic series
which cannot be summed and cannot provide us with the exact result.
Of course, for half integer $\nu$ when the asymptotic series terminates,
an exact result is obtained from the replica trick. 

The exact answer for the resolvent in the microscopic region shows
a compact-noncompact dichotomy. This dichotomy is natural in the 
supersymmetric approach where the compact part of the Goldstone
manifold is associated with the fermion-fermion sector and the
non-compact part is associated with the boson-boson sector. In the
fermionic replica trick this dichotomy is not at all clear but might
be hidden in the $n\to 0$ limit which is given by  an integration
over a $1\times 1$ matrix and and an integral over an $(n-1)\times(n-1)$
matrix. This might be another hint that it is possible to go
beyond perturbation theory within the replica framework.

Finally, we hope to have convinced the reader that the supersymmetric
method is the only $\sigma-$model approach that can provide
us with rigorous exact results. Even the calculation of a
small number of terms in the asymptotic expansion 
within the replica approach requires a tremendous
effort in the case of broken replica symmetry.

\vspace*{1.0cm}
\noindent{{\bf Acknowledgments}}
\vskip 0.5cm

This work was partially supported by the US DOE grant
DE-FG-88ER40388.  
Dominique Toublan  and Poul Damgaard are thanked for useful discussions.
J.J.M.V. is grateful to the Institute for Nuclear Theory at
the University of Washington for its hospitality and partial support
during the completion of this work. The work of D.D. is supported
by FAPESP (Brazilian Agency).

\vskip 1cm
\noindent{\large\bf Appendix}
\vskip 0.5cm

We are interested in expectation values
of traces of powers of   $n\times n$ Hermitian matrices 
with  matrix elements distributed according to
the Gaussian Unitary Ensemble (GUE).
Such averages are given by
\be
\Omega_q(p_1,p_2,\cdots ,p_q)&=&
\langle {\rm Tr}H^{p_1}{\rm Tr}H^{p_2}\cdots {\rm Tr}H^{p_q}\rangle =
\frac{\int dH\, e^{-\frac 12 {\rm Tr}H^2}{\rm Tr}H^{p_1}
{\rm Tr}H^{p_2}
\cdots {\rm Tr}H^{p_q}}{\int dH\, e^{-\frac 12 {\rm Tr}H^2}}\nonumber\\
&=&\frac{\int_{-\infty}^{\infty} \prod_{i=1}^{n} \left( dx_i
e^{-\frac 12 x_i^2}\right) 
\prod_{i<j}( x_i - x_j )^2\, 
\sum_{k_1=1}^n x_{k_1}^{p_1}
\sum_{k_2=1}^n x_{k_2}^{p_2}\cdots\sum_{k_q=1}^n x_{k_q}^{p_q}} 
{\int_{-\infty}^{\infty} \prod_{i=1}^{n} \left( dx_i
e^{-\frac 12 x_i^2}\right) 
\prod_{i<j}( x_i - x_j )^2} \nonumber\\
&=&\langle \omega_{p_1} \cdots \omega_{p_q} \rangle \, , 
\ee
where $\omega_l\equiv {\rm Tr}H^l $. All
correlators $\Omega_q(p_1,\cdots ,p_q)$ can be 
calculated recursively starting from $\Omega_1(0)=n$.
The recursion relations are derived from integrals
of total derivatives (Schwinger-Dyson equations),
\be 
\int_{-\infty}^{\infty} \prod_{i=1}^{n}  dx_i
\sum_{k=1}^n\partial_k \left( x_k^{a+1}
\sum_{j_1=1}^nx_{j_1}^{r_1} \cdots \sum_{j_b=1}^nx_{j_b}^{r_b}\,
\Delta_n^2 \, e^{-\frac 12 \sum_{i=1}^n x_i^2}\right)=0 .
\label{sd}\ee
It is useful to keep in mind the identity 
\be 
\sum_{k=1}^n\partial_k \left( x_k^{a+1}\Delta_n^2 \right) 
= \Delta_n^2 \sum_{l=1}^a\omega_l \omega_{a-l} .
\ee
As a sample calculation let us derive  $\Omega_3(1,1,4)$.
Choosing $b=2,r_1=4,r_2=1 $ and $a=-1$  in 
(\ref{sd}) we find
\be
\Omega_3(1,1,4)= 4 \Omega_2(1,3) + n\, \Omega_1(4).
\ee
Choosing now $b=1,\,\,a=-1$ and $r_1=3$ or $r_1=1$, respectively,
 we get
\be 
\Omega_2(1,3)&=&3\Omega_1(2),\nonumber\\
\Omega_2(1,1)&=&n.
\ee 
Finally, from $b=0$ and $a=2$ or $a=0$, respectively, we deduce
\be
\Omega_1(2)\, &=&\,n^2,\non\\
\Omega_1(4)\, &=& \, 2n\Omega_1(2) + \Omega_2(1,1). 
\ee
This system of equations results in
\be
\Omega_3(1,1,4) \, = \, 2n^4 + 13n^2. 
\ee
~For some special correlators we 
can easily derive a general formula, e.g.,
choosing $b=2k,\,\, a=-1 $ and $r_1=r_2=\cdots =r_{2k}=1$
we find
\be
\Omega_{2k}(1,1,\cdots , 1)\, =\, n^k (2k-1)!!\,. \label{closed}
\ee
Clearly all $\Omega_q (p_1,\cdots , p_q)$ with 
$\sum_i p_i $ being odd vanish identically.
Besides (\ref{closed}) we have used the following moments
in the calculation of this paper,
\be
\begin{array}{ll}
\Omega_1(0)\, = \, n, \,  &   \\
\vspace*{0.2cm}
\Omega_1(2)\, = \, n^2,  &  \\
\vspace*{0.2cm}
\Omega_1(4)\, = \, 2n^3 + n, \, & \, \Omega_2(2,2)\, = \, n^4 +2n^2, \\
\Omega_2(1,3) \, = \, 3n^2, \, & \, \Omega_3(1,1,2) \, = \, n^3 +2n, \\
\vspace*{0.2cm}
\Omega_1(6)\, =\, 5n^4+10n^2,\, &\,\Omega_2(2,4)\, =\, 2n^5 + 9n^3 +4n,\\
\Omega_2(3,3) \, = \, 12n^3 + 3n,\, & \,\Omega_2(1,5)\, = \, 10n^3 + 5n,\\
\Omega_3(1,1,4) \, = \, 2n^4 +13n^2,\, & \,
\vspace*{0.2cm}
\Omega_4(1,1,1,3)\, = \,9n^3 + 6n,\\
\Omega_1(8)\, = \, 14n^5 + 70n^3 + 21n,\, & \,
\Omega_2(2,6)\, = \, 5n^6 + 40n^4 + 60n^2,  \\
\Omega_2(4,4)\, = \, 4n^6 + 40n^4 + 61n^2, \,&  \,
\Omega_3(1,1,6)\, = \, 5n^5 + 70n^3 + 30n,  \\
\Omega_3(1,3,4)\, = \, 6n^5 + 75n^3 + 24n, \, & \, 
\vspace*{0.2cm}
\Omega_5(1,1,1,1,4)\,=\, 6n^5 + 75n^3 + 24n, \,\\ 
\Omega_{3}(2,4,4)\, = \, 4n^8+72n^6 + 381n^4 + 488n^2, \,&  \\ 
\Omega_{2}(4,6)\, = \, 10n^7+169n^5 + 610n^3 + 156n, &  \\
\Omega_{4}(1,1,4,4)\, = \, 4n^7+88n^5 + 661n^3 + 192n, &  \\
\Omega_{3}(4,4,4)\, = \, 8n^9+228n^7 + 2202n^5 + 6517n^3 + 1440n. & 
\end{array}
\ee
The sum of the coefficients of the correlators can always 
be checked by means of the $n=1$ case where all correlators
reduce to one dimensional  Gaussian integrals,
\be
\Omega_q(p_1,\cdots ,p_q)\vert_{n=1} \, = \, \left(\sum_i p_i\, 
-\, 1\right)!!. 
\ee
The coefficient of the highest order power in $n$ of 
the correlators   $\Omega_1(2k) $
can be checked
as follows. First, in the limit 
$n\to\infty $ one can solve the loop equation 
(or Virasoro constraints) which, for 
our Gaussian potential, gives the equation
\be 
\langle \, \sum_{i=1}^n \frac 1{p-x_i}\, \rangle \, = \,
\frac {p - \sqrt{p^2-4n}}{2},
\ee
where $p$ is a positive number assumed to be large 
(outside the semi-circle). After expanding
both sides of this equation in powers of $1/p$ we find
\be 
\lim_{n\to\infty} \Omega_1(2k) = \frac{(2n)^{k+1}(2k-1)!!}{2(k+1)!}.
\ee
%to check the same coefficient of
%other correlators we use the large $n$ factorization
%of observables , e.g. , $\lim_{n\to\infty}
%\Omega_2(p,q)=(\lim_{n\to\infty}\Omega_1(p))(\lim_{n\to\infty}\Omega_1(q)) $ 

\end{document}